\documentclass[aps,prb,showpacs,superscriptaddress]{revtex4}
\usepackage{graphicx}
\begin{document}

\title{First Principles Calculations of Ionic Vibrational Frequencies in
$\mathrm{PbMg_{1/3}Nb_{2/3}O_3}$ }

\author{S. A. Prosandeev$^{1,2}$, E. Cockayne$^1$, and E. P. Burton}
\affiliation{Ceramics Division, Materials Science and Engineering
Laboratory, National Institute of Standards and Technology,
Gaithersburg, Maryland 20899-8520; $^2$Physics Department, Rostov
State University, 5 Zorge St., 344090 Rostov on Don, Russia}
\begin{abstract}
Lattice dynamics for several ordered supercells with composition
$\mathrm{PbMg_{1/3}Nb_{2/3}O_3}$~ (PMN) were calculated with
first-principles frozen phonon methods.  Nominal symmetries of the
supercells studied are reduced by lattice instabilities. Lattice
modes corresponding to these instabilities, equilibrium ionic
positions, and infrared (IR) reflectivity spectra are reported.
\end{abstract} \maketitle

There is ample evidence of 1:1 (NaCl-type; the ``random layer
model" \cite{Davies}) short-range order (SRO) in PMN
\cite{Krause},  but first principles (FP) calculations with
sufficiently large supercells to realistically approximate a
SRO-disordered PMN crystal are prohibitively time consuming.
Relatively small supercells that might reasonably approximation
include the $[001]_{NCC'}$~ structure which was predicted to be
the PMN cation-ordering ground state (GS) \cite{Burton02}.

One objective of this study is to compute the relaxed GS for
different small supercells consisting of 15 and 30 ions, to
compare their energies, to compute the relaxed coordinates,
dynamical charges, vibrational frequencies, and infrared (IR)
reflection spectra for the GS. These results are important to
understand possible contribution of the ordered structures to IR
reflection and Raman spectra.

A second objective is understanding the nature of the soft
vibrational modes in the ordered structures of PMN. It has been
reported that relaxors exhibit both ferroelectric (FE) and
antiferroelectric (AFE) characteristics, \cite{Egami} and that
competition between FE and AFE fluctuations are the cause of
glass-type properties in PMN.

\begin{figure}
  \includegraphics[height=.4\textheight]{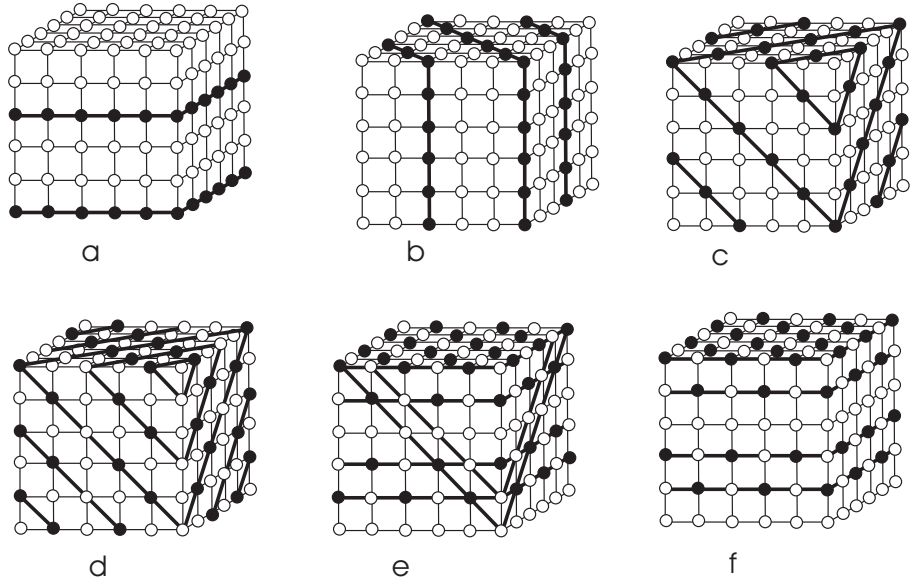}
  \caption{PMN structures considered in the present study:
[001]$_{NNM}$~(a), [110]$_{NNM}$~(b), [111]$_{NNM}$~(c),
[111]$_{NT}$~(d), [001]$_{NCC''}$~(e), and [001]$_{NCC'}$~ (f)}
\label{structures}
\end{figure}

\begin{figure}
  \includegraphics[height=.4\textheight]{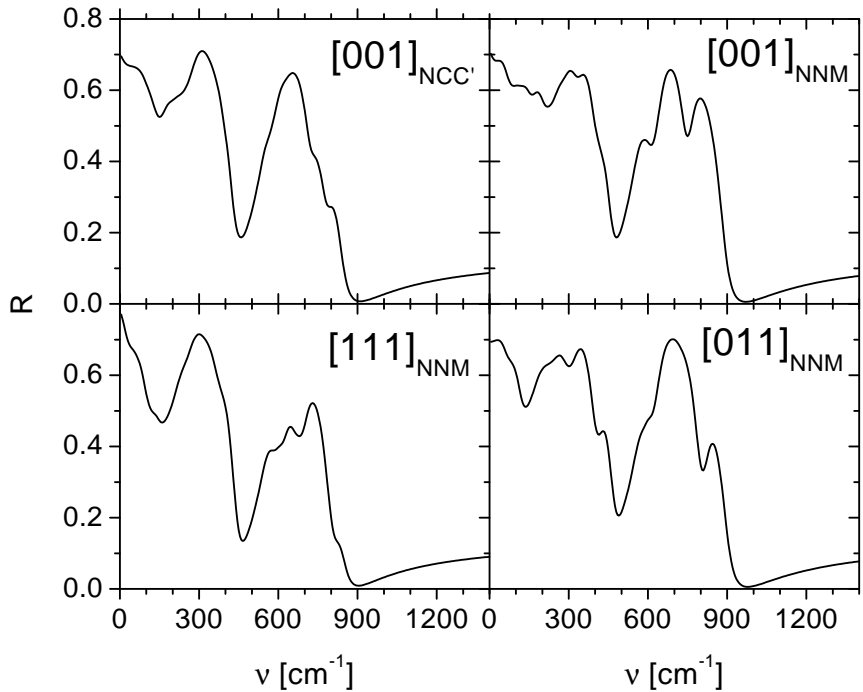}
  \caption{Computed reflection spectra for different ordered
structures of PMN} \label{reflection}
\end{figure}

All FP calculations were done with the Vienna {\it ab initio}
simulation package (VASP) \cite{Kresse1}. Several supercells of
PMN composition were considered (Fig. \ref{structures}). Our FP
computations show that all these structures are dynamically
unstable when the ions are placed on ideal perovskite positions,
and full relaxation often leads to suprisingly low symmetry. For
example, the FP GS of the $[001]_{NNM}$~ structure (an
$[001]_{2:1}$ superlattice) is monoclinic with the Pb close to Mg
displaced in the (0.18\AA,0.05\AA,-0.05\AA) and
(0.18\AA,0.05\AA,0.05\AA) directions. The Pb ions which are
between the Nb planes are mostly displaced in the $x$~ direction
by 0.27 \AA).

\begin{table}[!htbp]
 \caption{Ionic coordinates (in \AA) in the [001]$_{NCC'}$~ structure}
\begin{tabular}{c|ccc|c|ccc|c|ccc}
type& $x$  & $y$ & $z$ & type& $x$ & $y$ & $z$ & type& $x$ & $y$ &
$z$
\\ \hline
  Mg$_1$&    7.98& -0.02&  8.16&
  Mg$_2$&    4.00& -0.00&  4.08&
  Nb$_1$&    4.01& -0.09&  8.11\\
  Nb$_2$&    3.95& -0.04&  0.08&
  Nb$_3$&    7.93& -0.02&  0.00&
  Nb$_4$&    3.99& -3.98&  4.12\\
  O$_1$&     2.04&  0.40&  8.10&
  O$_2$&     5.96& -0.31&  8.04&\
  O$_3$&     4.40&  1.98&  8.12\\
  O$_4$&     3.66& -1.98&  8.10&\
  O$_5$&     2.05&  0.17&  0.00&
  O$_6$&     6.03& -0.03& 12.02\\
  O$_7$&     4.18&  2.01& 12.12&
  O$_8$&     4.00& -1.98& 12.07&
  O$_9$&     1.98&  0.00&  3.86\\
  O$_{10}$&  6.06&  0.06&  3.96&
  O$_{11}$&  4.02&  2.06&  3.87&
  O$_{12}$&  4.06& -2.02&  3.92\\
  O$_{13}$&  0.27& -0.01& 10.21&
  O$_{14}$&  4.00& -3.92&  1.94&
  O$_{15}$&  0.11& -0.04&  6.09\\
  O$_{16}$&  3.97&  0.14& 10.11&
  O$_{17}$&  4.12&  0.02&  1.95&
  O$_{18}$&  3.95&  0.20&  6.13\\
  Pb$_1$&    5.75&  1.99& 10.07&
  Pb$_2$&    5.96&  1.98&  2.41&
  Pb$_3 $&   5.84&  1.87&  6.30\\
  Pb$_4$&    1.89&  1.74& 10.10&
  Pb$_5$&    5.95& -2.04&  2.40&
  Pb$_6$&    2.06&  1.83&  6.27
 \end{tabular} \label{poscar30} \end{table}

Tetragonal $[001]_{NCC'}$~ PMN is also dynamically unstable, and
has has a wide spectrum of instabilities that are associated with
FE, AFE, octahedral tilting and other modes.  We relaxed the
$[001]_{NCC'}$~ structure after a random initial perturbation of
the ions. Relaxed coordinates are listed in Table \ref{poscar30}.
The basis vectors are: (4.00, -4.00, 0.00), (4.01,4.01,0.01),
(0.00,-0.01,3$\cdot$4.07) relative to cubic perovskite.
Relaxations of ionic coordinates are a complex mixture of
octahedral deformations around Mg-ions (with rather large frozen
angles) plus AFE, and FE, Pb-displacements in opposition to
neighboring O-ions. The $x$-direction displacements differ from
$y$-direction displacements. Pb-ions have either two, or four, Mg
nearest neighbors (nn): Pb-ions with two Mg nn are displaced, by
0.25 \AA, towards the centroid of the nearest Mg-Mg second-nn
pair. Pb-ions with four Mg-nn are not displaced. Although we can
not prove that the structure in Table \ref{poscar30} is the PMN
$[001]_{NCC'}$ GS (further relaxations may still reduce its energy
in the case), it is dynamically stable and has lower energy than
PMN $[001]_{NCC'}$~ that is relaxed with only FE displacements, or
only octahedral tilting.

Dynamical charges $Z^{\star}$~ for the tetragonal [001]$_{NCC'}$~
structure are listed in Table \ref{Berry30}. Nb ions that have Nb
nn in the $\pm  \alpha$ directions have particularly large
$Z^{\star}_{\alpha\alpha}$.  Dynamical charges for Pb also exhibit
significant anisotropy, and environment dependence.

\begin{table}[!htbp]
 \caption{The dynamical charges of ions in the 30-ion tetragonal supercell
of PMN}
\begin{tabular}{c|ccc|c|ccc|c|ccc}
ion $i$ & $Z^*_{izz}$ & $Z^*_{ixx}$ & $Z^*_{iyy}$ &ion $i$ &
$Z^*_{izz}$ & $Z^*_{ixx}$ & $Z^*_{iyy}$&ion & $Z^*_{izz}$ &
$Z^*_{ixx}$ & $Z^*_{iyy}$ \\ \hline
 Mg$_1$  & 2.74 & 2.65 & 2.65 &
 Mg$_2$  & 2.74 & 2.65 & 2.65 &
 Nb$_1$  & 6.48 & 6.00 & 6.00 \\
 Nb$_2$  & 7.87 & 9.11 & 9.11 &
 Nb$_3$  & 7.87 & 9.11 & 9.11 &
 Nb$_4$  & 6.48 & 6.00 & 6.00 \\
 O$_1$   &-2.61 &-3.62 &-2.82 &
 O$_2$   &-2.61 &-3.62 &-2.82 &
 O$_3$   &-2.61 &-2.82 &-3.62 \\
 O$_4$   &-2.61 &-2.82 &-3.62 &
 O$_5$   &-1.99 &-7.04 &-2.09 &
 O$_6$   &-1.99 &-7.04 &-2.09 \\
 O$_7$   &-1.99 &-2.09 &-7.04 &
 O$_8$   &-1.99 &-2.09 &-7.04 &
 O$_9$   &-2.61 &-3.62 &-2.82 \\
 O$_{10}$&-2.61 &-3.62 &-2.82 &
 O$_{11}$&-2.61 &-2.82 &-3.62 &
 O$_{12}$&-2.61 &-3.62 &-2.82 \\
 O$_{13}$&-4.15 &-2.57 &-2.57 &
 O$_{14}$&-5.71 &-2.42 &-2.42 &
 O$_{15}$&-3.93 &-2.44 &-2.44 \\
 O$_{16}$&-5.71 &-2.42 &-2.42 &
 O$_{17}$&-4.15 &-2.57 &-2.57 &
 O$_{18}$&-3.93 &-2.44 &-2.44 \\
 Pb$_1$  & 3.52 & 4.41 & 4.41 &
 Pb$_2$  & 3.52 & 4.41 & 4.41 &
 Pb$_3$  & 4.17 & 3.93 & 3.93 \\
 Pb$_4$  & 3.52 & 4.41 & 4.41 &
 Pb$_5$  & 3.52 & 4.41 & 4.41 &
 Pb$_6$  & 4.17 & 3.93 & 3.93
 \end{tabular} \label{Berry30} \end{table}

According to Ghosez \cite{Ghosez}, dynamical (Born) charges $Z^*$~
are associated with the average field $E_{av}$~ but not with the
local field $E_l$, as the Szigeti charge $Z$~ is \cite{Szigeti}.
However the product of the Born charge with the average field
$Z^*E_{av}$~ must equal the product of the Szigeti charge with the
local field, $ZE_l$~ \cite{Ghosez}. It follows that

\begin{equation}
Z^*_i=\frac{Z_i E_i}{E_{av}}
\end{equation}
Consider the average field in a stack of $xy$-layers with
different dielectric permittivities $\varepsilon_1$~ and
$\varepsilon_2$:

\begin{equation}
E_{av}=DL\left(\frac{d_1}{\varepsilon_1}+\frac{d_2}{\varepsilon_2}
\right)
\end{equation}
where $D$~ is an external field, $L$~ is sample length parallel to
the field, $d_1$~ and $d_2$~ are the lengths of stacked layers.
The local field in a uniform dielectric is

\begin{equation}
E_i=\frac{D}{\varepsilon_i}
\end{equation}
 Substituting these expressions into the
initial one yields:

\begin{equation}
Z^*_i=\frac{Z_i}{\varepsilon_i}
\left[\frac{d_1}{\varepsilon_1}+\frac{d_2}{\varepsilon_2} \right]
\end{equation}
It follows that, with two layers, one with low and the other with
high electronic dielectric permittivity, $Z^*_i$~ in the more
polarizable layer decreases, and $Z^*_i$~ in the less polarizable
layer increases. The FP calculations follow this trend:
$Z^*_{Mg}$, which is less polarizable, increases while $Z^*_{Nb}$,
which is more polarizable, decreases. Within the Nb plane, one
calculates more normal values for $Z^*_{Nb}$.

\begin{table}[!htbp]
 \caption{The diagonal frequencies of the dynamical matrix for the 30-ion
 $[001]_{NCC'}$~ supercell of PMN}
\begin{tabular}{c|ccc|c|ccc|c|ccc}
ion & $z$ & $x$ & $y$ &ion & $z$ & $x$ & $y$&ion & $z$ & $x$ & $y$
\\ \hline

 Mg$_1$  &346 & 328 & 340 &
 Mg$_2$  &276 & 340 & 344 &
 Nb$_1$  &279 & 271 & 290 \\
 Nb$_2$  &322 & 276 & 270 &
 Nb$_3$  &318 & 286 & 251 &
 Nb$_4$  &301 & 288 & 283 \\
 O$_1$   &385 & 537 & 333 &
 O$_2$   &262 & 608 & 251 &
 O$_3$   &399 & 331 & 521 \\
 O$_4$   &256 & 267 & 648 &
 O$_5$   &300 & 640 & 248 &
 O$_6$   &263 & 657 & 257 \\
 O$_7$   &297 & 272 & 625 &
 O$_8$   &281 & 246 & 633 &
 O$_9$   &292 & 577 & 302 \\
 O$_{10}$ & 281 & 612 & 278 &
 O$_{11}$ & 285 & 278 & 600 &
 O$_{12}$ & 279 & 299 & 613 \\
 O$_{13}$ & 555 & 264 & 294 &
 O$_{14}$ & 661 & 267 & 261 &
 O$_{15}$ & 735 & 214 & 207 \\
 O$_{16}$ & 557 & 241 & 250 &
 O$_{17}$ & 733 & 233 & 253 &
 O$_{18}$ & 596 & 244 & 245 \\
 Pb$_1$  & 108 & 88  & 76  &
 Pb$_2$  & 81  & 78  & 81  &
 Pb$_3$  & 93  & 81  & 68  \\
 Pb$_4$  & 105 & 76  & 89  &
 Pb$_5$  & 88  & 79  & 70  &
 Pb$_6$  & 92  & 67  & 76
 \end{tabular} \label{diag30}\end{table}

For the 30-ion $[001]_{NCC'}$~ structure, computed diagonal
frequencies of the dynamical matrix are listed in Table
\ref{diag30}. The frequencies of Nb-, Mg- and bending O-vibrations
are all between (200 to 350) cm$^{-1}$. The O-vibrations along
Mg-O and Nb-O bonds range from (520 to 735) cm$^{-1}$. The Pb
diagonal frequencies are in the range (67 to 108) cm$^{-1}$.

As in experiment \cite{Reaney}, computed IR reflection spectra
(Fig.  \ref{reflection}) consist of three main reststrahlen bands
of the vibrational modes that are typical of perovskites: the
first group is below 120 cm$^{-1}$; the second spreads from 150
cm$^{-1}$~ to 400 cm$^{-1}$; and the third is from 500 cm$^{-1}$~
to 800 cm$^{-1}$. The two lower bands split into two subbands
each. Assignments of these bands can be made on the basis of the
diagonal frequencies shown in Table \ref{diag30}. In experimental
data \cite{Reaney}, these groups of lines are rather compact as in
the computed [001]$_{NCC'}$~ and [111]$_{MNN}$~ structures.
However the experimentally determined magnitude of the
reflectivity in the interval from 500 cm$^{-1}$~ to 800 cm$^{-1}$~
is lower than in the computation. This could be connected with an
overestimation of the oxygen dynamical charge and/or with large
damping for some frequencies in this interval (an estimated
damping constant of 60 cm$^{-1}$~ was used for all frequencies),
or it could be that the systems studied here are not sufficiently
representative of SRO-disordered PMN.

The lowest calculated optical frequency in the equilibrium 30-ion
[001]$_{NCC'}$~ structure is 24 cm$^{-1}$. It is lower than the
lowest Pb diagonal frequency (60 cm$^{-1}$) shown in Table
\ref{diag30} because of the interaction among the ionic
vibrations. This mode is a mixture of opposite but not fully
compensated Pb-displacements. Other Pb-related modes are spread
over the interval (24 to 129) cm$^{-1}$. Ferroelectric Pb
displacements are at (40 and 60 to 90) cm$^{-1}$~ although some
contribution to $\varepsilon$~ exists in the whole interval from
24 cm$^{-1}$~ to 129 cm$^{-1}$. Some modes in this interval have
large Pb components for definite sites that implies that these
modes are quasilocal. The same can be said of some modes in the
frequency interval from 500 cm$^{-1}$~ to 800 cm$^{-1}$: some of
these modes have a very large components corresponding to oxygen
vibrations along Nb-O-Mg bonds.

Raman spectra show broad lines with gradual temperature dependence
in a wide temperature interval \cite{Yuzuk,Husson}. The presence
of these lines would be forbidden if the ions were in symmetric
environments. Ionic displacements due to disorder and symmetry
breaking can explain the existence of these Raman lines. A
possible measure of the intensities of these lines is the square
of the projection of the ionic displacements, from symmetric
positions, onto the vibrational modes: $S_i = \left| \left< {\bf
v}_d| {\bf v}_i \right> \right|^2$, where ${\bf v}_d$~ is the
vector of the frozen displacements, and ${\bf v}_i$~ is the vector
of the $i$-th vibration in the displacements' representation.

Low-frequency Raman lines (about 50 cm$^{-1}$~ may be related to
the quasilocal Pb-O vibrations \cite{Husson}. Note that as is
typical of Pb-based perovskites that also have large B-cations,
such as Mg, that Pb-vibrational branches have relatively small
dispersion and are unstable across most of the Brillouin zone. The
particular instabilities that freeze in depend sensitively on the
local electric fields produced by the Nb-Sc configuration.
Freezing of lattice instabilities creates low-symmetry Pb-sites,
and also allows AFE Pb-vibrations to couple with FE vibrations.

Calculated structural instabilities in the various supercells
imply reductions in their formation energies relative to the
values reported in Burton and Cockayne \cite{Burton02}, but the
hierarchy of formation energies and predicted GS, $[001]_{NCC'}$,
are the same.  Column one in Table \ref{Energy}  corresponds to
the structures with symmetry restrictions imposed (for instance,
tetrgonal symmetry was imposed for [001]$_{MNN}$~ and
[001]$_{NCC'}$). Column two includes relaxation energy, to the GS
for each structure.

%in each case different symmetry restrictions were imposed Sergey.

There are many possible ordered derivatives of the random layer
model\cite{Davies} that have (111) Nb-layers which alternate with
(111) (Nb$_{1/3}$Mg$_{2/3}$)-layers in which Nb and Mg are
ordered; e.g.  Fig.  \ref{structures}d. Some mixed layers in this
structure are ordered in stripes and others in an Nb-honeycomb
pattern with Mg's at hexagon centers.  This structure has a large
formation energy relative to the others in Figs. \ref{structures}.
Apparently intralayer cation ordering increases the total energy
relative to a disordered configuration. This is consistent with
the conclusion of Hoatson et al.\cite{Hoatson} who obtained an
improved inverse Monte Carlo fit to NMR data with the assumption
of Mg-Mg and Nb-Nb clustering within mixed layers. The structure
shown in Fig. \ref{structures}e also has higher energy than
$[001]_{NCC'}$.

\begin{table}[!htbp]
 \caption{The supercell's energy (in eV per 30-ion supercell)}
\begin{tabular}{c|cc}
& ideal structure & relaxed structure \\ \hline
 $[001]_{NNM}$ & 0.825 & 0.583 \\
 $[110]_{NNM}$ & 0.710 & 0.155 \\
 $[111]_{NNM}$ & 0.696 & 0.150 \\
 $[001]_{NCC'}$& 0.523 &0 \\
 \end{tabular} \label{Energy} \end{table}

Finally, our FP computations have shown that the relaxation of the
ionic coordinated in the small supercells of PMN does not change
the hierarchy of the energy of these structures: [001]$_{NCC'}$~
remains to be the GS among the small supercells considered. The
low frequency vibrations in this structure is connected with AFE
Pb vibrations and with Pb-O stretching mode. The computed IR
reflection spectrum, qualitatively, corresponds to experimental
data \cite{Reaney} although there are some discrepancies in the
line magnitudes at high frequencies.

S.A.P. appreciates discussions with Toulouse, Svitelskiy, Petzelt,
Kamba and Yuzuk.


\begin{thebibliography}{999}
\bibitem{Davies}P. K. Davies and M. A. Akbas, J. Phys. Chem. Sol.
{\bf 61}, 159 (2000).
\bibitem{Krause} H.B. Krause, J.M. Cowley and J. Wheatley, Acta. Cryst.
{\bf A35}, 1015 (1979).
\bibitem{Burton02}B. P. Burton and E. Cockayne, Ferroelectrics
{\bf 270}, 173 (2002).
\bibitem{Egami}T. Egami, Ferroelectrics {\bf 267}, 101 (2002).
\bibitem{Kresse1} G. Kresse and J. Hafner, Phys. Rev. {\bf B47},
558 (1993).
\bibitem{Ghosez}Ph. Ghosez, Phys. Rev.  {\bf B58}, 6224 (1998).
\bibitem{Szigeti}B. Szigeti, Trans. Faraday Soc. {\bf 45}, 155 (1949).
\bibitem{Reaney}I. M. Reaney, J. Petzelt, V. V. Voitsekhovskii, F.
Chu, and N. Setter, J. Appl. Phys. {\bf 76}, 2086 (1994).
\bibitem{Yuzuk}V. I. Torgashev, Yu. I. Yuzyuk, L. T. Latush,
P. N. Timonin, and R. Farhi, Ferroelectrics {\bf 199}, 197 (1997).
\bibitem{Husson}E. Husson, L. Abello, and A. Morell, Mat. Res.
Bull. {\bf 25}, 539 (1990).
\bibitem{Hoatson} G.L. Hoatson, D.H. Zhou, F. Fayon, D. Massiot, and R.L.
Vold
Phys. Rev. {\bf B66} 224103 (2002).


\end{thebibliography}
\end{document}